\documentclass{article}
\usepackage{spconf,amsmath,graphicx,subcaption,amssymb}
\usepackage{enumitem}
\usepackage{soul}

\def\x{{\mathbf x}}

\def\blank{\left<\text{b}\right>}
\def\y{{\mathbf y}}
\def\yhat{{\mathbf{\hat{y}}}}

\title{RNN-T Models Fail to Generalize to Out-of-Domain Audio: Causes and Solutions}
\name{\begin{tabular}{c}Chung-Cheng Chiu, Arun Narayanan, Wei Han, Rohit Prabhavalkar, Yu Zhang, Navdeep Jaitly$^{1}$\thanks{$^{1,2}$Work conducted while the authors were at Google},\\ Ruoming Pang, Tara N. Sainath, Patrick Nguyen$^{2}$, Liangliang Cao, Yonghui Wu\end{tabular}}
\address{Google Inc, D. E. Shaw Group$^{1}$}

\begin{document}

\maketitle

\begin{abstract}
In recent years, all-neural end-to-end approaches have obtained state-of-the-art results on several challenging automatic speech recognition (ASR) tasks.  
However, most existing works focus on building ASR models where train and test data are drawn from the same domain. This results in poor generalization characteristics on mismatched-domains: e.g., end-to-end models trained on short segments perform poorly when evaluated on longer utterances. In this work, we analyze the generalization properties of streaming and non-streaming recurrent neural network transducer (RNN-T) based end-to-end models in order to identify model components that negatively affect generalization performance.
We propose two solutions: combining multiple regularization techniques during training, and using dynamic overlapping inference.
On a long-form YouTube test set, when the non-streaming RNN-T model is trained with shorter segments of data, the proposed combination improves word error rate (WER) from $22.3\%$ to $14.8\%$; when the streaming RNN-T model trained on short \emph{Search} queries, the proposed techniques improve WER on the YouTube set from $67.0\%$ to $25.3\%$. 
Finally, when trained on Librispeech, we find that dynamic overlapping inference improves WER on YouTube from $99.8\%$ to $33.0\%$.
\end{abstract}


\section{Introduction}
\label{sec:intro}
The last decade has seen rapid improvements in automatic speech recognition (ASR) technology 
through advances in deep learning~\cite{HintonDengYuDahlEtAl12}.
Recently, there has been growing interest in building so-called \emph{end-to-end} ASR systems -- systems consisting of a single neural network, which directly output character-based or word-based units: e.g., connectionist temporal classification (CTC)~\cite{GravesFernandezGomezSchmidhuber06,graves2014towards} with character~\cite{HannunCaseCasperCatanzaroEtAl14, AmodeiAnathanarayananAnubhaiBaiEtAl16} or word~\cite{soltau2016neural, AudhkhasiRamabhadranSaonPichenyEtAl17} targets; attention-based encoder-decoder models~\cite{ChanJaitlyLeVinyals16, ChorowskiBahdanauSerdyukChoEtAl15, WatanabeHoriKimHersheyEtAl17, Chiu2018state, Chiu2018}; and the recurrent neural network transducer (RNN-T)~\cite{Graves2013, RaoSakPrabhavalkar17, Tara2020}.

Most previous works investigating end-to-end models have evaluated models in the setting where training and test utterances are relatively short (i.e., tens of seconds) and drawn from the same domain\footnote{In this context, we use `\emph{domain}' to refer to utterances which share a common property. E.g., \emph{audiobooks} (long read speech utterances~\cite{PanayotovChenPoveyKhudanpur15}), or voice search queries (short utterances).}~\cite{Chiu2018state,  ParkChanZhangChiuEtAl19, LuscherBeckIrieKitzaEtAl19}.
In previous work, we identified two problems that affect end-to-end models: first, we observe that end-to-end models are particularly sensitive to a \emph{domain-mismatch} between training and inference, caused by overfitting to the training domain~\cite{Narayanan2019longform}. Since end-to-end models learn all components jointly, the effect is more pronounced than would be expected in conventional models~\cite{Narayanan2018}.
A second problem -- a specific kind of domain mismatch -- is the observation that end-to-end models trained on short training segments 
do not perform well when decoding much longer utterances during inference (e.g., longer YouTube videos)~\cite{Narayanan2019longform,Chiu19longform}; this problem is particularly acute for non-streaming attention-based models~\cite{Chiu19longform}, but, somewhat surprisingly, also affects streaming end-to-end models such as RNN-T.\footnote{By streaming models, we refer to models which produce and update hypotheses for each input speech frame (e.g., the CTC, or RNN-T models with unidirectional encoders). Similarly, we refer to models which examine all of the input speech before producing an output hypothesis (e.g., RNN-T with a bi-directional encoder, or attention-based encoder-decoder models) as non-streaming models.}

Our previous works proposed a number of solutions to address these problems:
training on diverse domains~\cite{Narayanan2019longform}; simulating long-form speech by manipulating the encoder/decoder states~\cite{Narayanan2019longform}; or by performing inference over short overlapping segments which can be assembled into the complete hypothesis~\cite{Chiu19longform}.
Although our proposed solutions improved performance on out-of-domain and long-form audio, our previous works did not characterize the \emph{fundamental reasons} for the degradation in performance.
In the present work, we perform a detailed analysis of the RNN-T model to determine which models components are primarily responsible for this performance degradation, finding that the \emph{encoder} network in the model is most susceptible to overfitting.
In light of this observation, we reinterpret previously proposed solutions~\cite{Narayanan2019longform, Chiu19longform} as additional regularization constraints imposed on the model to prevent overfitting, and find that combining multiple regularization techniques results in the best performance.
In experimental evaluations, we decode a YouTube test set using three RNN-T models: a model trained using short-segments of YouTube data; a model trained using short-segments of Search data; and a model trained on the Librispeech dataset.
We find that combining various regularization techniques improves the models trained on YouTube and Search data by $33.6\%$ and $62\%$,  respectively.
In combination with our proposed dynamic overlapping inference technique (See Section~\ref{sec:overlap}), our mismatched Librispeech trained models show dramatic word error rate (WER) improvements from $99.8\%$ to $33.0\%$ on the YouTube test set.

\section{RNN Transducer}
\label{sec:rnnt}
The RNN-T model was proposed by Graves~\cite{Graves12, Graves2013} as an improvement over CTC~\cite{GravesFernandezGomezSchmidhuber06}. As with CTC, the RNN-T model introduces a special blank symbol, $\blank$, which models the alignments between the speech frames, $\x = [\x_1, \cdots, \x_T]$, and the output label sequence, $\y = [y_1, \cdots, y_U]$.
We denote the number of speech frames by $T$, with each $\x_t \in \mathbb{R}^d$, and $\mathcal{Y}$ denotes the set of output labels with $y_u \in \mathcal{Y}$.
The set of all valid frame-level alignments, can be written as: $B(\x, \y) = \left\{\yhat = (\hat{y}_1, \cdots, \hat{y}_{T+U})\right\}$, where $\hat{y}_i \in \mathcal{Y} \cup \left\{\blank\right\}$, such that $\yhat$ is \emph{identical} to $\y$ after removing all blank symbols. During training, RNN-T uses the forward-backward algorithm to maximize $P(\y|\x)$, taking all valid alignments into consideration.

\begin{figure}
  \centering
  \includegraphics[width=\columnwidth]{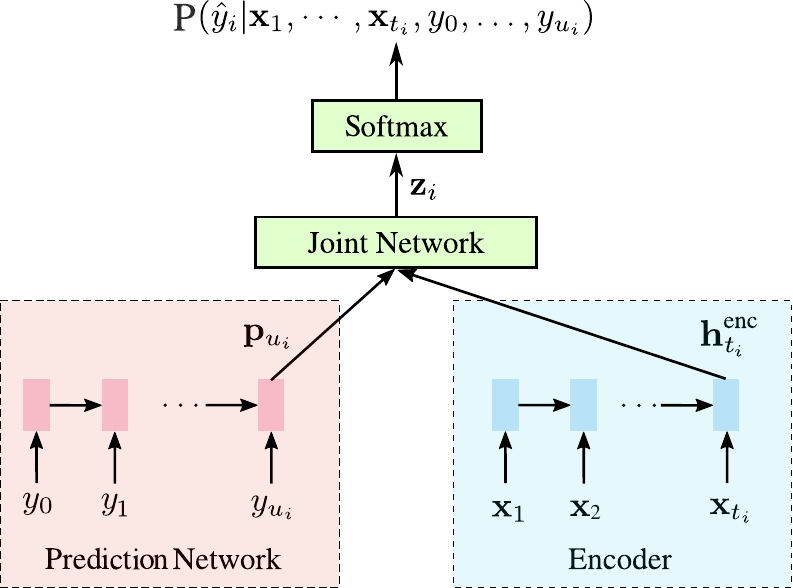}
  \caption{Block diagram of an RNN-T model~\cite{Graves12, Graves2013}.}
  \label{fig:rnnt}
\end{figure}

The RNN-T model, depicted in Fig.~\ref{fig:rnnt}, consists of an \emph{encoder}, a \emph{prediction network} (an LSTM~\cite{HochreiterSchmidhuber97} network), and the \emph{joint network} (a feed-forward network) which integrates information from the other two. More implementation details can be found in~\cite{Narayanan2019longform}.

\section{The Generalization Problem}
\label{sec:issue}
This section characterizes the generalization problem for streaming and non-streaming models and presents experimental observations that identify components that contribute to poor generalization.

\subsection{Non-streaming ASR Models}

Our experimental setup is similar to~\cite{Chiu19longform}. The training data is extracted from YouTube videos ~\cite{Soltau2017}. The training utterances are generally short: the $50^{th}$ percentile length is $5.13$ seconds and the $90^{th}$ percentile is $12.67$ seconds.  During training we filter out utterances that are longer than $15.36$ seconds.  We evaluate the model on two YouTube test sets, \textit{YT-short} and \textit{YT-long}.  \textit{YT-short} is comprised of $119$ videos with length ranging from $2$ to $10$ minutes, with a total duration of $11.37$ hours. 
\textit{YT-long} is comprised of $87$ videos with length ranging from $41.8$ seconds to $30$ minutes, with a total duration of $24.12$ hours.  The videos in both test sets are much longer than the training samples, thus allowing us to test the long-form generalization of non-streaming ASR models.

Our RNN-T model's encoder stacks a macro layer $3$ times, where the macro layer consists of $1$-D convolution with filter width $5$ and $512$ filters with stride $1$, a $1$-D max pooling layer with width $2$ and stride $2$, and $3$ bidirectional LSTM layers with $512$ hidden units in each direction and a $1,536$-dimensional projection layer~\cite{Pang2018}.  The prediction network has a unidirectional LSTM with $1,024$ hidden units.  The output network has $512$ hidden units and the final output uses a $4$k word piece model \cite{SchusterNakajima12}.  As input, the model uses $80$-dimensional log-Mel features, computed with a $25$ms window, shifted every $10$ms.

\begin{figure}[t]
    \centering
    \includegraphics[width=\columnwidth]{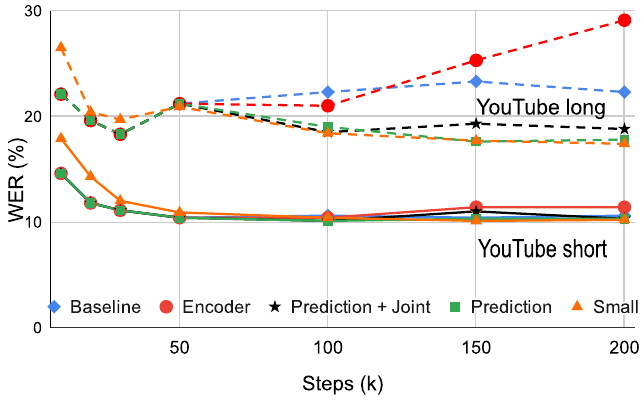}
    \caption{WERs for non-streaming model on \textit{YT-short} (solid line) and \textit{YT-long} (dotted line) as a function of training steps.
    Results are also shown when only the encoder, prediction, or the prediction + joint networks are trained after $50$k steps of training all components.}
    \label{fig:yt}
\end{figure}

\noindent \textbf{Observations: } As shown in Figure~\ref{fig:yt}, the RNN-T model trained with short utterances exhibits high WER, mainly due to deletion errors, when evaluated on both test sets.

\begin{figure}[t]
    \centering
    \includegraphics[width=\columnwidth]{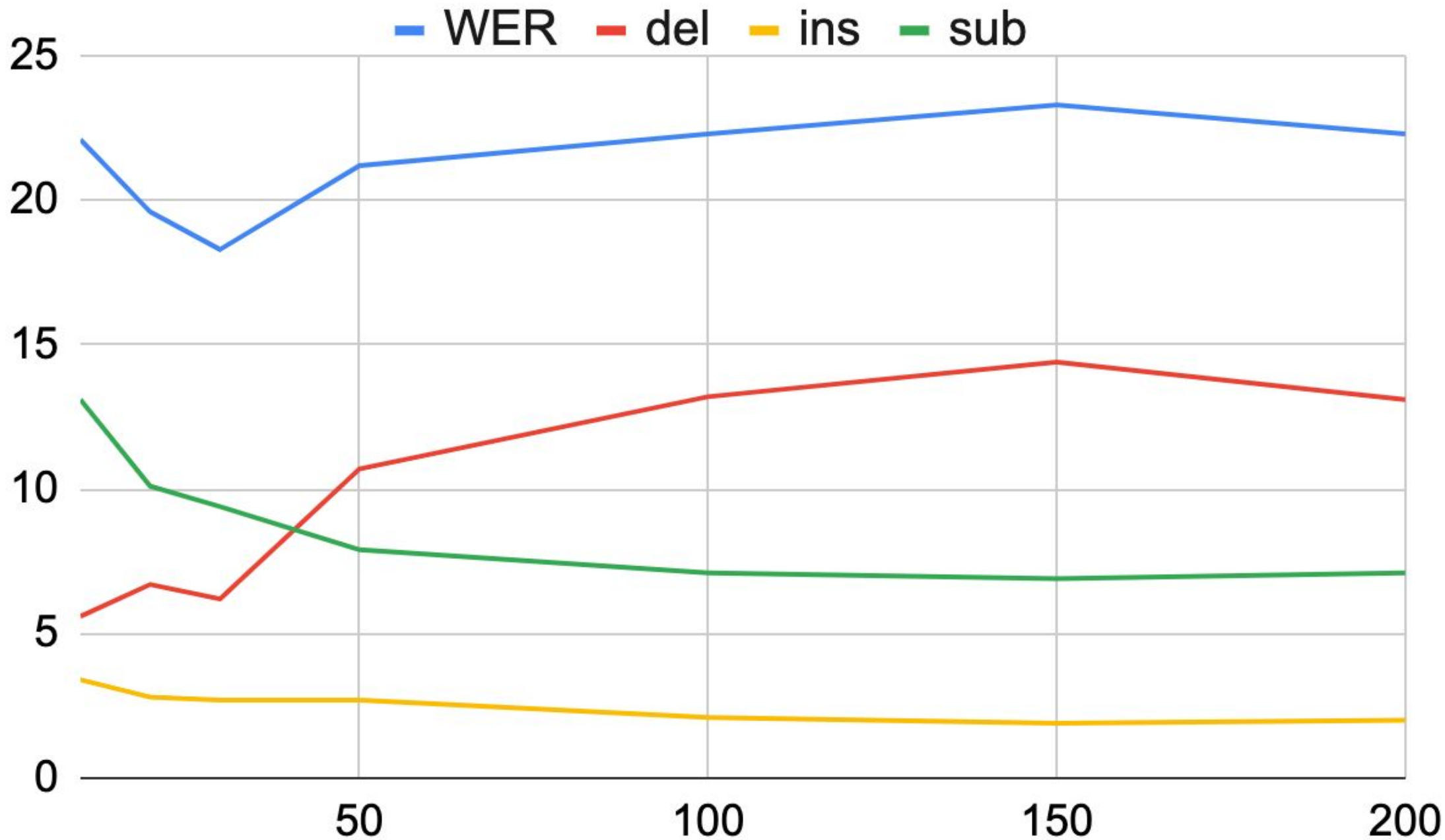}
    \caption{WERs and the respective deletion, insertion, and substitution errors for non-streaming model on \textit{YT-long} as a function of training steps.}
    \label{fig:ytlongform}
\end{figure}

Analyzing overall word errors as a function of training steps, we observe that the model makes more deletion errors as training proceeds; the phenomenon is particularly significant on \textit{YT-long}, which has longer utterances. The behavior is illustrated in Figure~\ref{fig:ytlongform} with WERs of the model as a function of training steps. The deletion errors grow as training proceeds, which indicates a correlation between overfitting and the underlying long-form issues.

To better understand this overfitting issue we compare various training setups that freeze parts of the model after initially training all components for $50$k steps: updating only the encoder; or the prediction network; or both the prediction and the joint networks.
As can be observed in Fig.~\ref{fig:yt}, high word error rates (WERs) are correlated with models that update the encoder. Models with a smaller encoder, or the ones that do not update the encoder layers show better generalization than the baseline model. Furthermore, updating only the encoder results in worse performance on long-form sets than the baseline model, which indicates that the overfitting issue does not simply reflect the number of parameters being updated but is particularly associated with the encoder.

\begin{figure*}[t]
    \centering
    \includegraphics[width=\textwidth]{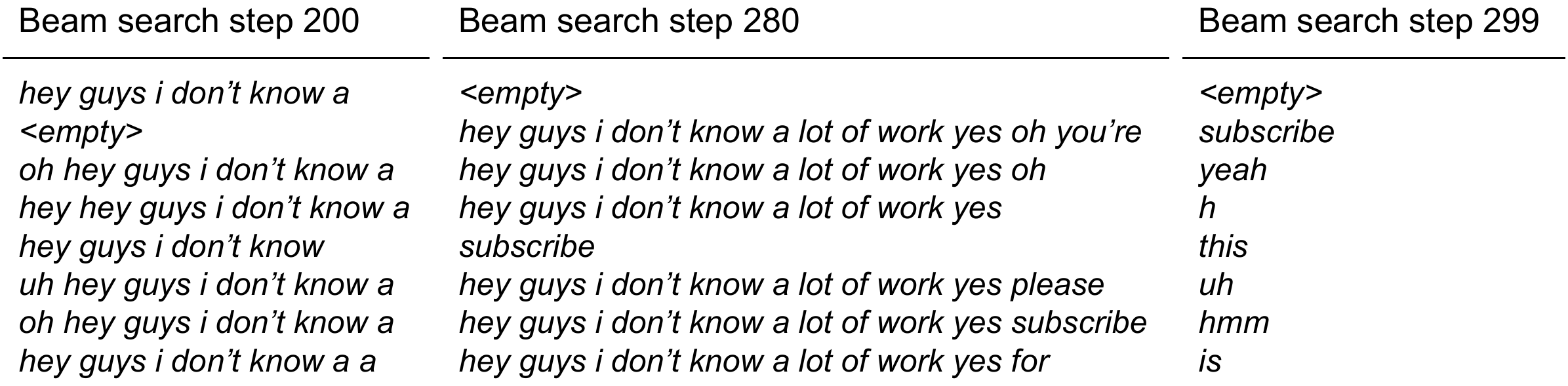}
    \caption{An example of beam search steps for one of the utterance in \textit{YT-long} that exhibit high deletion errors. The model is able to recognize the utterance at the beginning, but as the beam search proceeds the $<$\textit{empty}$>$ hypothesis starts to dominate and eventually fill the beam with hypotheses branched from the $<$\textit{empty}$>$ hypothesis. Once the beam is filled with these hypotheses, the beam search process can no longer recover and result a final hypothesis with high deletion errors.}
    \label{fig:beam_search}
\end{figure*}

The higher word error rate is caused mainly because of significantly higher deletion errors in a few utterances; the deletion errors are not as prominent in a lot of the utterances.  A further analysis on model's beam search step shows these high deletion errors are related to unstable predictions of $<$blank$>$ symbols.  Fig.~\ref{fig:beam_search} shows an example of beam search failure on one such utterance with high deletion errors. 
In this example, the $<$\textit{empty}$>$ hypothesis stays in the beam during the initial stage of beam search, along with other valid alternatives. But eventually, it forces these alternatives out of the beam as hypotheses conditioned on $<$\textit{empty}$>$ hypothesis starts to outscore the valid ones. In the end, the beam is filled with short, frequent phrases, unrelated to the acoustic content of the utterance. One interesting thing about this example is that the word ``subscribe'' never appears in the utterance, but the model still hypothesizes the word, most likely due to its high frequency in the training data.

\begin{figure}[t]
    \includegraphics[width=\columnwidth]{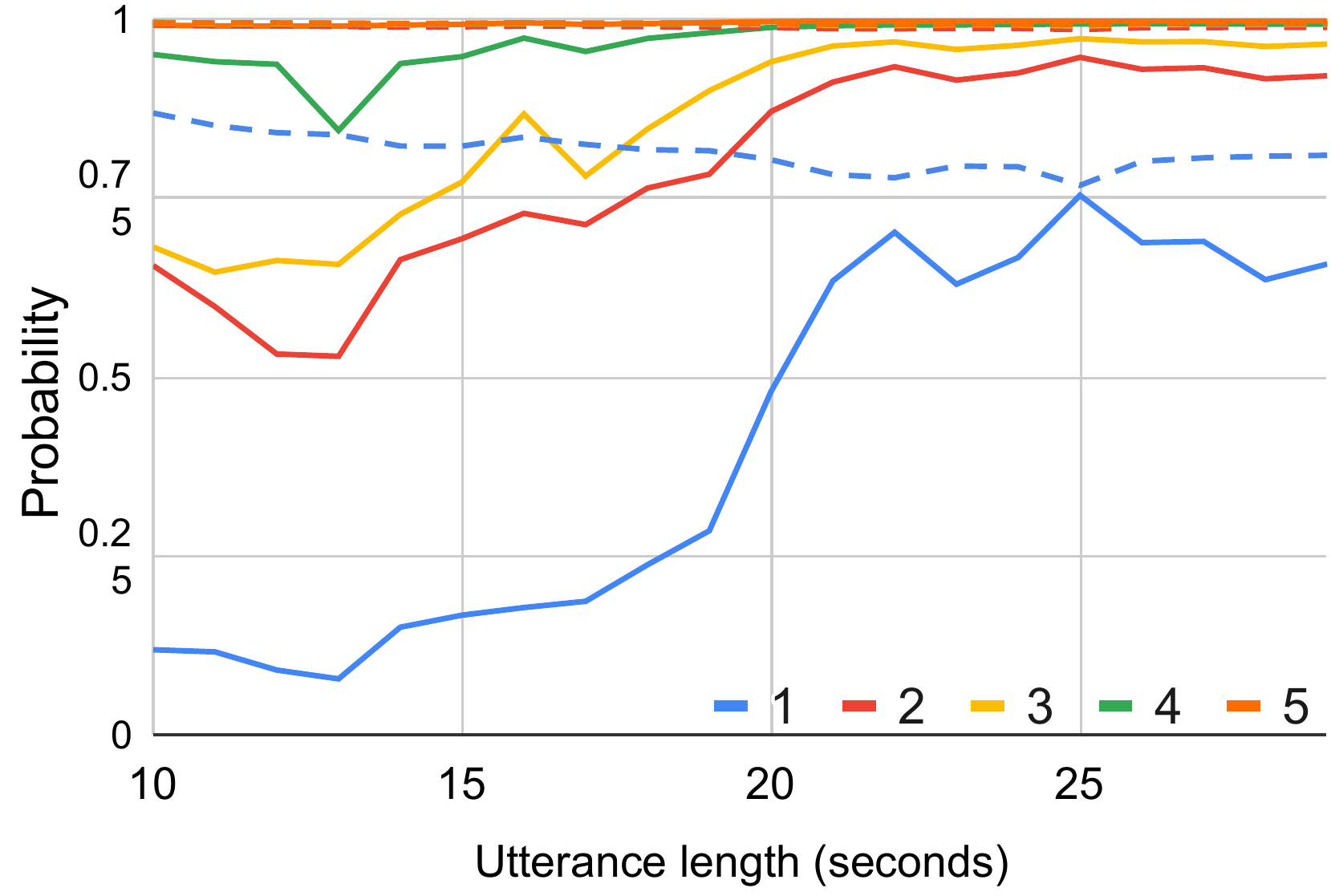}
    \caption{Probability of blank symbol for steps $1$ -- $5$, as a function of utterance length.
    Dotted and solid lines correspond to predictions at $50$k and $200$k steps, respectively.
    At $200$ steps the model exhibited inconsistent prediction as a function of utterance length.}
    \label{fig:ytblankprob}
\end{figure}

To analyze the correlation between encoder overfitting and unstable predictions for blank symbols, we sample one video with very high deletion errors from \textit{YT-long}, and segment the first $10, 11, \dotsc, 29$ seconds of audio. Typically, for these YouTube tasks, the model converges at around $200k$ steps. And the deletion errors phenomenon starts to emerge at around $50k$ steps. Therefore, we then evaluate the model at $50$k and $200$k steps on these segments and compare the probability of blank prediction for the first $5$ steps.
Note that the encoder is bidirectional, so the entire audio segment will influence the predictions for the first $5$ steps.  
Furthermore, the first $5$ frames contain no speech and the models should, ideally, predict blanks with high probability. 
As can be seen in Fig.~\ref{fig:ytblankprob}, at $50$k steps, the model has similar confidence amongst different utterance lengths.
On the other hand, at $200$k steps the model's confidence varies a lot with respect to the utterance length, in particular for utterances longer than $15$ seconds. 
Note that during training the model has only seen utterances less than $15.36$ seconds long.
Thus, as training proceeds, the encoder's prediction for blank symbols fails to generalize well on longer utterances.  
Since this model has bidirectional LSTMs, the backward LSTM contributes towards the high variance of blank probability at the first few steps of predictions.
This affects the model's hypotheses in multiple ways: first, a high blank probability would result in partial hypotheses consisting of a sequence of  blank tokens to have a higher probability than the correct sequence; 
eventually this blank sequence would also cause other partial hypotheses that have fewer blanks and are more accurate to be dropped from the beam, causing additional search errors. This results in the WER being dominated by deletions.

\subsection{Streaming ASR Models}
A typical streaming application is voice search on mobile phones. We, therefore, choose this task for the streaming use-case. Due to latency constraints, the model size is smaller than the non-streaming case; our experimental setup mimics those in \cite{Narayanan2019longform}. 128-dimensional log-mel features from 4 contiguous frames are stacked to form a 512 dimensional input, which is then subsampled by a factor of 3 along the time dimension. 
The RNN-T model has 8 encoder layers made up of unidirectional LSTMs. Each layer has 2048 units and a projection layer with 640 outputs units \cite{prabhavalkar2016compression}. The decoder consists of 2 unidirectional LSTMs, also with 2048 units and 640 projections similar to the encoder layers. The joint network has a single layer with 640 units. The target is represented by a sequence of word piece tokens \cite{SchusterNakajima12}, with a vocabulary size of 4096.

The training data consists of anonymized and hand-transcribed utterances representative of the Google search traffic \cite{Narayanan2019longform}. We use multicondition training (MTR) to simulate noisy conditions, and randomly downsample the data from 16~kHz to 8~kHz to improve generalization to varying input sample rates.
The training utterances are short, with mean and median duration of $6.3$ and $4.8$ seconds, respectively. The $90^{th}$ percentile is $10.9$ seconds. The $50^{th}$ and $90^{th}$ percentiles for the  target sequences of word-pieces, are respectively, $5$ and $17$ tokens. 
As test sets, we use a mix of in-domain and out-domain data.
A test set similar to the training domain and composed of $60$ hours of anonymized and hand transcribed search queries forms the in-domain test set (\textit{Search}; median length is $6$ seconds).
Our out-of-domain test set consists of 7 hours of speech generated using a text-to-speech system~\cite{gonzalvo2016recent}, which is acoustically simple but much longer than the training utterances (\textit{TTS-Audiobook}; median length is $62$ seconds).

\begin{figure}[t]
    \includegraphics[width=1.0\columnwidth]{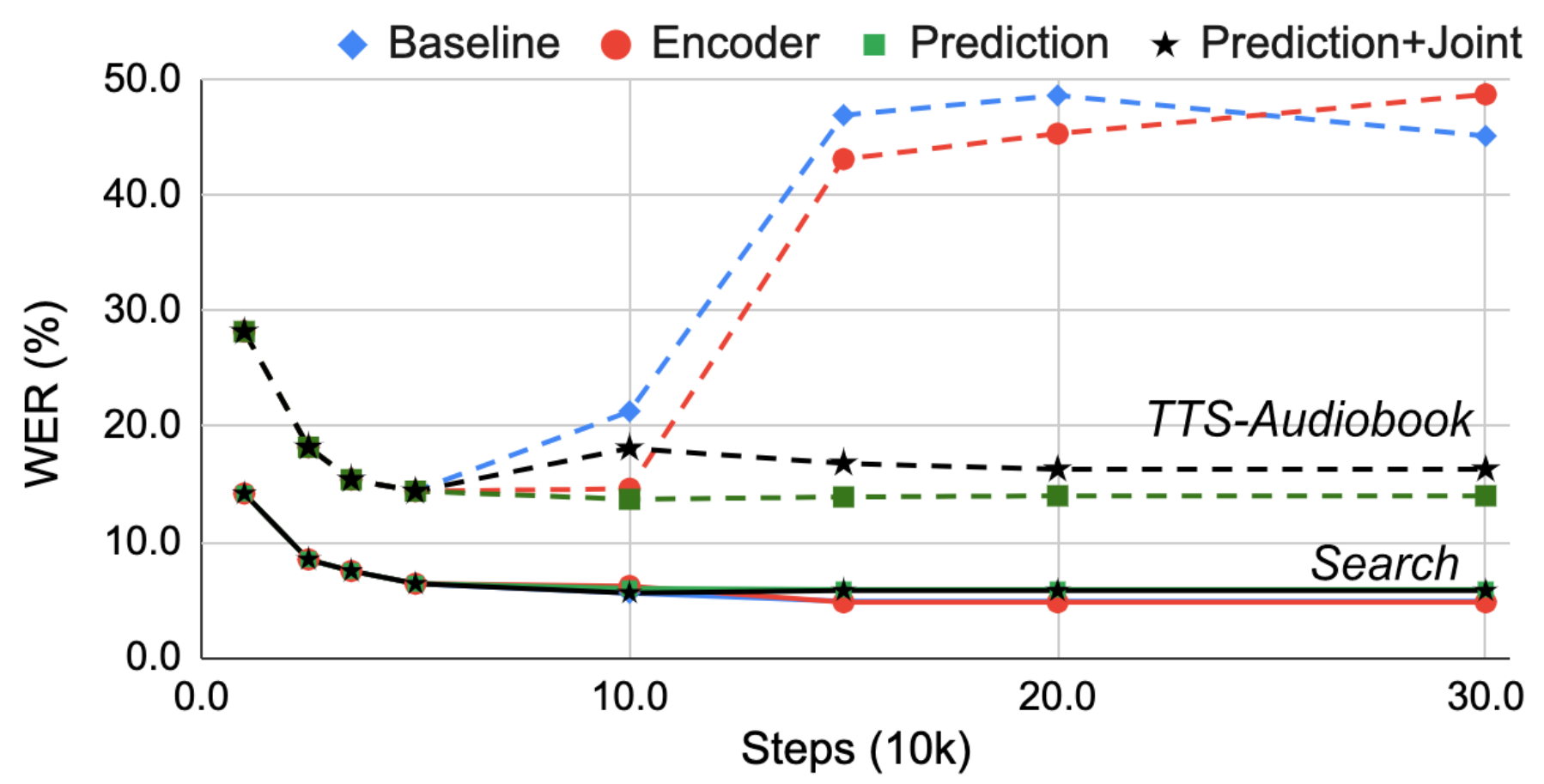}
    \caption{Word error rates on in-domain (\textit{Search}) and out-of-domain (\textit{TTS-Audiobook}) test sets as a function of training steps. On \textit{Search} (solid line), all models improve WER with more training; on \textit{TTS-Audiobook}, only the models that do not train the encoder further generalize well.}
    \label{fig:seq-issues}
\end{figure}

\noindent \textbf{Observations: } Similar to the non-streaming models, we observe that the streaming model also overfits to the training domain after approximately $50$k steps as shown in Fig.~\ref{fig:seq-issues}. Unlike non-streaming models, however, performance keeps improving on the in-domain \textit{Search} set with more training. Performance on the \textit{TTS-Audiobook} set, in contrast, gets worse as training progresses. Next, we freeze parts of the model after $50$k steps, as  before, and continue training just the encoder; the prediction network; or both the prediction and joint networks. Confirming the observations made for the non-streaming models, freezing the encoder layers and only updating the prediction, or the prediction and the joint layers prevents this overfitting behavior. After $300$k steps, both the baseline, which updates all model parameters, and the model that only updates encoder layers obtain WERs of $45.1$ and $48.7$, respectively, on \textit{TTS-Audiobook};
the model that only updates the prediction network obtains a WER of $14.0\%$.
Thus, the degradation in performance is significantly less severe when the model does not update the encoder after $50$k steps.

The results presented in the streaming and the non-streaming case indicate that the encoder is most responsible for the overfitting behavior of RNN-T. In the next section, we explore various regularization strategies to reduce overfitting.

\section{Regularization Cocktail}
\label{sec:method}
As the generalization issue is caused by 
encoder overfitting, it can be remedied effectively through regularization. 
Different domains and architectures can benefit from different regularization techniques, and thus we 
combine them
during training to create a \emph{regularization cocktail}:
\begin{itemize}[leftmargin=*]
    \item \textbf{Variational Weight Noise:} Variational weight noise adds Gaussian noise to the weight matrix during training~\cite{Graves2011}, and has been shown to be effective in improving generalization~\cite{Graves2013}.  In our approach we start the training process without noise, and start adding it after a predefined number of steps.  The weight noise is re-sampled at every training step.
    \item \textbf{SpecAugment: } SpecAugment~\cite{ParkChanZhangChiuEtAl19,specaugment2020} is a data augmentation algorithm that alters the spectrogram of the input utterances.  The approach applies time warping, time masking, and frequency masking on the spectrogram, and trains the model to be robust to such data augmentation.
    \item \textbf{Random state sampling and random state passing: } Random state sampling (RSS) and random state passing (RSP) were proposed in \cite{Narayanan2019longform} as a way to address generalization of streaming RNN\-T models to long-form speech. RSS assumes that LSTM states follow a normal distribution and samples initial LSTM states from it during training. RSS is readily applicable for bidirectional models as well. RSP, on the other hand, saves LSTM states from each mini-batch during training, and uses them as initial states for examples in the subsequent batch. When used with unidirectional models, this mimics random concatenation of examples during training.
\end{itemize}

\section{Dynamic Overlapping Inference}
\label{sec:overlap}
\begin{figure}[t]
    \includegraphics[width=\columnwidth]{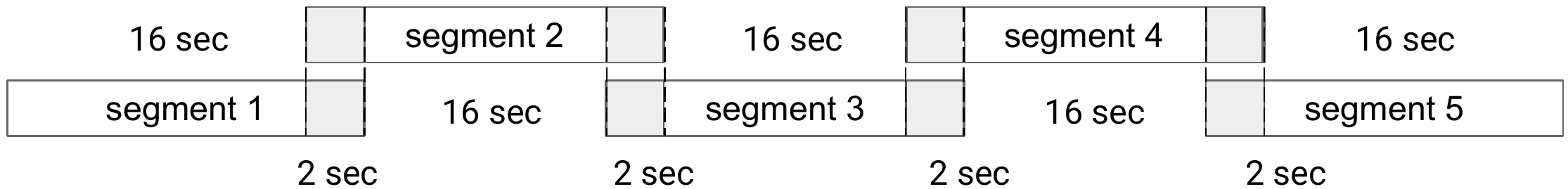}
    \caption{Dynamic overlapping inference with 16 second segments and 2 second overlap. Reducing the overlapped regions from 8 seconds as in~\cite{Chiu19longform} greatly improves inference efficiency.}
    \label{fig:doi}
\end{figure}
In addition to improving generalization during training, we attempt to improve generalization during decoding with overlapping inference~\cite{Chiu19longform}.
This method segments a long utterance into multiple fixed-length segments which are decoded independently.
Since each segment lacks context from neighboring segments, we allow some overlap between successive segments, and merge the decoded hypotheses in the overlapped region. The original method~\cite{Chiu19longform} was proposed in the context of models which do not have any alignment information for the hypothesis, and required a 50\% overlap between segments and $2\times$ the computational cost compared to regular inference.

Here, we extend overlapping inference to relax the 50\% overlap requirement.
Our proposed algorithm -- dynamic overlapping inference (DOI) -- infers frame-level alignment (alignment, here, refers to the frame associated with each non-blank label) obtained from each RNN-T hypothesis, to match and merge hypotheses between segments. More specifically, we can only match words that have been aligned into the overlapping region between consecutive segments. On unmatched pairs of words with matching alignment, we adopt a tie-breaking method similar to \cite{Chiu19longform} by assigning confidence scores based on the relative position of the word in a segment, and favoring those that are closer to segment centers.
The process is illustrated in Fig~\ref{fig:doi}.

Compared to \cite{Chiu19longform}, dynamic overlapping inference allows us to use a significantly shorter overlap interval, thus greatly increasing computational efficiency while maintaining the same decoding quality.

\section{Experiments}
\label{sec:exp}
We evaluate generalization performance of the proposed techniques for the models described in Sec.~\ref{sec:issue}. The experiment setup is identical to those in Sec.~\ref{sec:issue}. All models are implemented with Lingvo~\cite{lingvo}.

\begin{table}[t]
  \begin{center}
  \begin{tabular}{l|cccccc}
    Models & \multicolumn{2}{c}{YT-short} & \multicolumn{2}{c}{YT-long} & \multicolumn{2}{c}{Call-center}\\
    {} & Reg. & DOI & Reg. & DOI & Reg. & DOI\\ \hline
    Base & $10.6$ & $9.7 $ & $22.3$ & $17.0$ & $27.6$ & $22.4$\\
    SpecAugment & $9.4$ & $9.4$ & $15.9$ & $15.3$ & $21.5$ & $20.3$\\
    {} + RSS & $9.3$ & $9.2$ & $15.6$ & $15.2$ & $19.6$ & $19.5$\\
    {} + RSS + VN & $\textbf{9.1}$ & $\textbf{9.0}$ & $\textbf{14.8}$ & $\textbf{14.9}$ & $\textbf{19.3}$ & $\textbf{19.2}$\\\hline
  \end{tabular}
  \end{center}
  \caption{Non-streaming model WERs with regular inference (Reg.) and DOI on \textit{YT-short}, \textit{YT-long} and \textit{Call-center} test sets. VN applies variational weight noise with standard deviation $0.05$ on the encoder. SpecAugment uses $10$ dynamic time masks up to $4\%$ of audio length, and $2$ frequency masks up to $27$ dims.
  }
  \label{table:yt-exp}
\end{table}

\subsection{Non-Streaming Models}

The results using the regularization cocktail and dynamic overlapping inference (DOI) are shown in Tab.~\ref{table:yt-exp}. All regularization techniques and their combinations help improve performance. In particular, SpecAugment + RSS + VN obtains a $14.2\%$ improvement on \textit{YT-short} and a $33.6\%$ improvement on \textit{YT-long}, and DOI obtains $8.5\%$ ($10.6$ to $9.7$) and $23.8\%$ ($22.3$ to $17$) improvement on \textit{YT-short} and \textit{YT-long} respectively.  We further evaluate the model on a long-form call-center test set described in \cite{Narayanan2019longform} to assess its robustness on unseen domain. The proposed regularization cocktail improves WER by $30.1\%$ when using regular inference, and improves WER by $18.8\%$ ($27.6$ to $22.4$) when using DOI. In general, DOI provides significant improvement when models have generalization issue on the target domain, and provide similar quality as regular inference for models that do no have this issue.

\begin{table}[t]
  \begin{center}
  \begin{tabular}{l|ccc}
    Models        & Search & TTS-Audiobook & YT-short \\ \hline
    Baseline      & $4.9$ & $48.6$ & $67.0$ \\
    VN            & $4.7$ & $31.3$ & $59.8$ \\
    SpecAugment   & $\textbf{4.6}$ & $16.5$ & $52.9$ \\
    {} + RSP      & $5.1$ & $\textbf{11.9}$ & $27.3$ \\
    {} + RSP + VN & $5.1$ & $\textbf{11.9}$ & $\textbf{25.3}$ \\\hline
  \end{tabular}
  \end{center}
  \caption{WERs using streaming RNN-T models trained on \textit{Search} data. VN applies variational noise with standard deviation $0.03$ on all layers. SpecAugment uses $2$ time masks with widths up to $1.5$ seconds and $2$ frequency masks up to $27$ dimensions.}
  \label{table:seq-exp}
\end{table}

\subsection{Streaming Models}

Results are shown in Tab.~\ref{table:seq-exp}. As with the non-streaming models, the model with multiple regularizations gave the best improvements. SpecAugment$+$VN$+$RSP obtains a 76\% improvement on \textit{TTS-Audiobook} and a 62\% improvement on \textit{YT-short}. Other combinations also help, but are slightly worse than SpecAugment$+$VN$+$RSP.
It should be noted that some of these models still perform better at $50$k checkpoint. For example, SpecAug$+$VN obtains 14.5\% and 22.5\% on \textit{TTS-Audiobook} and \textit{YT-short}, respectively, at $50$k steps. Although the combination doesn't completely prevent overfitting, the degradation, as the model converges on the training data, is much lower than the baseline.
We note that DOI does not help with streaming models, likely because it relies on alignment and end-to-end streaming models are know to produce poor alignments unless they are constrained during training \cite{senior2015acoustic}.

\begin{table}[t]
  \begin{center}
  \begin{tabular}{c|cc}
     & Reg. & DOI\\ \hline
    Test & $3.2$ ($0.2/0.4/2.6$) & $3.2$ ($0.2/0.4/2.6$)\\
    Test Other & $7.8$ ($0.7/0.8/6.3$) & $7.8$ ($0.6/0.9/6.3$)\\
    YT-short & $99.8$ ($99.5/0.1/0.2$) & $33.0$ ($3.6/7.2/22.2$)\\\hline
  \end{tabular}
  \end{center}
  \caption{WERs (deletions/insertions/substitutions) for the RNN-T model trained on Librispeech, with regular inference (Reg.) and DOI. Test and Test Other refers to Librispeech Test and Librispeech Test Other sets respectively. Here DOI uses $16$s window with $2$s overlap.}\label{table:lb-exp}
\end{table}

\subsection{Librispeech}

The final set of results are when the RNN-T model is trained on Librispeech ~\cite{Librispeech}.  We follow the architecture of LAS-6-1280 described in~\cite{ParkChanZhangChiuEtAl19}. The prediction network has the same LSTM setup as the LAS decoder.  The joint network has $640$ hidden units, and uses the same word piece model.  The results are shown in Tab.~\ref{table:lb-exp}.  Despite achieving low WERs on the Librispeech test sets, the model exhibits high deletion errors on \textit{YT-short}.  DOI achieves the same WERs on the Librispeech test sets, and reduces the deletion error of \textit{YT-short} from $99.5\%$ to $3.6\%$. The model still has a WER of $33.0\%$ after using DOI, mainly due to high substitutions ($22.2\%$). Our further analysis found that the majority of the errors are caused by model producing phonetically similar words. This is likely caused by the limited vocabulary the Librispeech model is exposed to during training.

The Librispeech RNN-T model exhibited $>99\%$ WERs on \textit{YT-short} even at early stages of training, and, therefore, multiple regularizations do not remedy the issue as well as DOI.  The regularization cocktail mainly addresses generalization to new domains.  When the gap between training and test domains is large, addressing the long-form issue during inference provides a more robust solution.

\section{Conclusions}
\label{sec:conclusion}

This work presents an analysis of the generalization problem observed in RNN-T based end-to-end ASR models. 
Our results demonstrate that the model's affinity to  predict blank sequences when there is a mismatch between training and test distributions causes this problem, which results in high deletion rates. 
Our analysis identified the root cause of this problem to be encoder overfitting.
We proposed a regularization cocktail that significantly improves the performance of streaming and non-streaming RNN-T models trained with large-scale data.
For models trained on a smaller dataset, where regularization alone doesn't improve performance, we proposed a dynamic overlapping inference strategy that significantly improves generalization.
Future work will explore alternative model architectures and regularization techniques that address the generalization of models trained on smaller datasets.

\bibliographystyle{IEEEbib}
\bibliography{refs}

\end{document}